\documentclass[10pt, letterpaper]{article}

\usepackage{opex3}
\usepackage{cite}
\usepackage{amsmath}
\usepackage{graphicx}

\newcommand{\ud}{\mathrm{d}}
\newcommand{\eqn}[1]{(\ref{#1})}
\newcommand{\fig}[1]{Fig. \ref{#1}}
\newcommand{\Fig}[1]{Figure \ref{#1}}
\newcommand{\Figs}[1]{Figures \ref{#1}}
\newcommand{\figs}[1]{Figs. \ref{#1}}

\begin{document}

\title{Simultaneous Amplitude and Phase Measurement for Periodic Optical Signals Using Time-Resolved Optical Filtering}

\author{Keang-Po Ho$^{1,2}$, Hsi-Cheng Wang$^1$, Hau-Kai Chen$^1$, Cheng-Chen Wu$^1$}%

\address{$^1$Institute of Communication Engineering, $^2$Department of Electrical Engineering,
National Taiwan University, Taipei 10671, Taiwan}

\email{kpho@cc.ee.ntu.edu.tw}
\homepage{http://cc.ee.ntu.edu.tw/~kpho}

\begin{abstract}
Time-resolved optical filtering (TROF) measures the spectrogram or sonogram by a fast photodiode followed a tunable narrowband optical filter.
For periodic signal and to match the sonogram, numerical TROF algorithm is used to find the original complex electric field or equivalently both the amplitude and phase.
For phase-modulated optical signals, the TROF algorithm is initiated using the craters and ridges of the sonogram.
\end{abstract}



\section{Introduction}

Recently, differential phase-shift keying (DPSK) signal has received renewed interests for long-haul lightwave transmissions \cite{gnauck02, zhu03, rasmussen03, cai04, charlet04a, xu04, gnauck05, ho05}.
Compared with conventional on-off keying, DPSK signal provides 3-dB improvement in receiver sensitivity.
With its low peak power and  constant-intensity, DPSK signal can also provide better tolerance to fiber nonlinearities than on-off keying \cite{mizuochi03, xu04, gnauck05, ho05}.
However, unlike on-off keying signal, the signal phase or the complex signal cannot directly be measured using a photodiode.
Method to measure the phase and amplitude of an optical signal is important for the further development of DPSK signals.

Conventionally, the complex electric field is measured by a quadrature receiver with a $90^\circ$ optical hybrid \cite{walker84, hodgkinson85, ho05}.
Recently, optical phase was measured based on similar technique with a pulse source as local oscillator \cite{dorrer05}.
Those measurements are similar to a receiver that can find both in- and quadrature-phase components of an electric field \cite{taylor04, gagnon05}.
Requiring a local oscillator with about the same wavelength as the signal, those measurements may be difficult to conduct and not available in most communication laboratories.

A complex signal including the amplitude and phase is measured here using time-resolved optical filtering (TROF).
The method is the time-frequency duality of frequency-resolved optical grating (FROG) \cite{trebino93, kane99, trebino97}.
FROG measures the spectrogram of an optically sampled pulse of
\begin{equation}
I_\mathrm{FROG}(\omega, \tau) = \left| \int E(t) G(t - \tau) e^{-j \omega t} \ud t \right|^2,
\label{frogtrace}
\end{equation}
\noindent where $E(t)$ is the complex electric field, $G(t)$ is the waveform of the grating pulse, and $\tau$ is the time-delay of the gating pulse.
Numerically, the FROG trace is the calculation of the spectrogram using short-time Fourier transform \cite{cohen89, qian}.

TROF measures the sonogram of an optically filtered signal of
\begin{equation}
I_\mathrm{TROF}(t, \nu) = \left| \int E(\omega) H(\omega - \nu) e^{j \omega t} \ud \omega \right|^2,
\label{troftrace}
\end{equation}
\noindent where $E(\omega)$ is the spectrum of the optical signal, $H(\omega)$ is the frequency response of the tunable optical filter, and $\nu$ is the center frequency of $H(\omega)$.
In FROG, the spectrogram is used to retrieve the phase of $E(t)$.
In TROF, the sonogram is used to retrieve the phase of $E(\omega)$ and equivalently, via inverse Fourier transform, the phase of $E(t)$.

When the sonogram of \eqn{troftrace} is compared with the spectrogram of \eqn{frogtrace},  the exchange of frequency and time transforms the methods between FROG and TROF.
As a dual of FROG, the theory for FROG is applicable to TROF.
However, FROG is typically for short optical pulses \cite{kane99, trebino97, trebino93, baltuska98, nishizawa01, gallmann00}.
Theory for FROG for uniqueness  generally assumes a signal with finite support, i.e., time-limited short pulse \cite{trebino97, trebino}.
Here, we are interested of using TROF for continuous-time infinite-support periodic DPSK signals.
For the convenience to use fast Fourier transform, we assume a periodic optical signal in this paper.
Typical measurements can use the whole $2^7-1$ pseudo-random binary sequence (PRBS) as a periodic signal.

With the same application as FROG, sonogram was used to measure the complex electric field of an short optical pulse \cite{chilla91, taira01, reid99}.
Unlike previous measurement based on nonlinear process like two-photon absorption \cite{reid99} or optical sampling \cite{taira01}, electrical sampling is sufficient for typical 10- and 40-Gb/s signals.
Although the trace of \eqn{troftrace} is called sonogram for optical pulses, it is just another method to find the spectrogram of \eqn{frogtrace} \cite{spectrogram}.
The TROF trace of \eqn{troftrace} is by itself also a spectrogram.

The sonogram or spectrogram is commonly used in time-resolved chirp measurement of an optical signal and directly measured using a sampling scope followed a tunable optical filter \cite{linke85, agilent}.
However, time-resolved chirp measurement can use a filter bandwidth far larger than that for TROF.
To certain extend, a sonogram is another representation of the original data for time-resolved chirp measurement but for the calculation of the complex electric field.
This paper does not invent new measurement equipment but provides a new interpretation of existing measured data to obtain new results.

The remaining parts of this paper are the following: Sec. \ref{sec:dir} shows the TROF traces of some DPSK signals. 
The shape of the TROF traces can be used to initiate the algorithm to find the signal from TROF traces.
Sec. \ref{sec:inv} first discusses the equipment and method to measure the TROF traces and then the inverse problem to find the signal from the measured TROF trace.
A numerical optimization method is used.
Secs. \ref{sec:dis} and \ref{sec:end} are the discussion and conclusion of the paper, respectively. 

\section{TROF Trace of DPSK Signals}
\label{sec:dir}

For a periodic signal with a period of $T$ and expressed as a Fourier series of %
\begin{equation}
E(t) = \sum_k c_k  \exp \left( \frac{2 \pi j k t}{T} \right),
\label{ESignal}
\end{equation}
the sonogram or spectrogram of \eqn{troftrace} becomes
\begin{equation}
I_\mathrm{TROF}(t, \nu)
  = \left| \sum_k c_k H \left( \frac{2 \pi k}{T} - \nu \right) \exp \left( \frac{2 \pi j k t}{T} \right) \right|^2,
\label{trofperiod}
\end{equation}
\noindent where $c_k$ are the Fourier coefficients of $E(t)$, $H(\omega)$ is the frequency response of the tunable narrowband optical filter, and $\nu$ are the centered frequencies in the measurement of the TROF trace.

The goal of this paper is the ``inverse'' problem to find the signal of $E(t)$ using the TROF trace of \eqn{trofperiod}.
For illustration purpose and to understand the problem, the ``direct'' problem is simulated.
The ``direct'' problem finds the TROF trace of a known signal, in here, a phase-modulated optical signal.

\begin{figure}[htbp]
\begin{center}
\begin{tabular}{cc}
\includegraphics[width = 0.45 \textwidth]{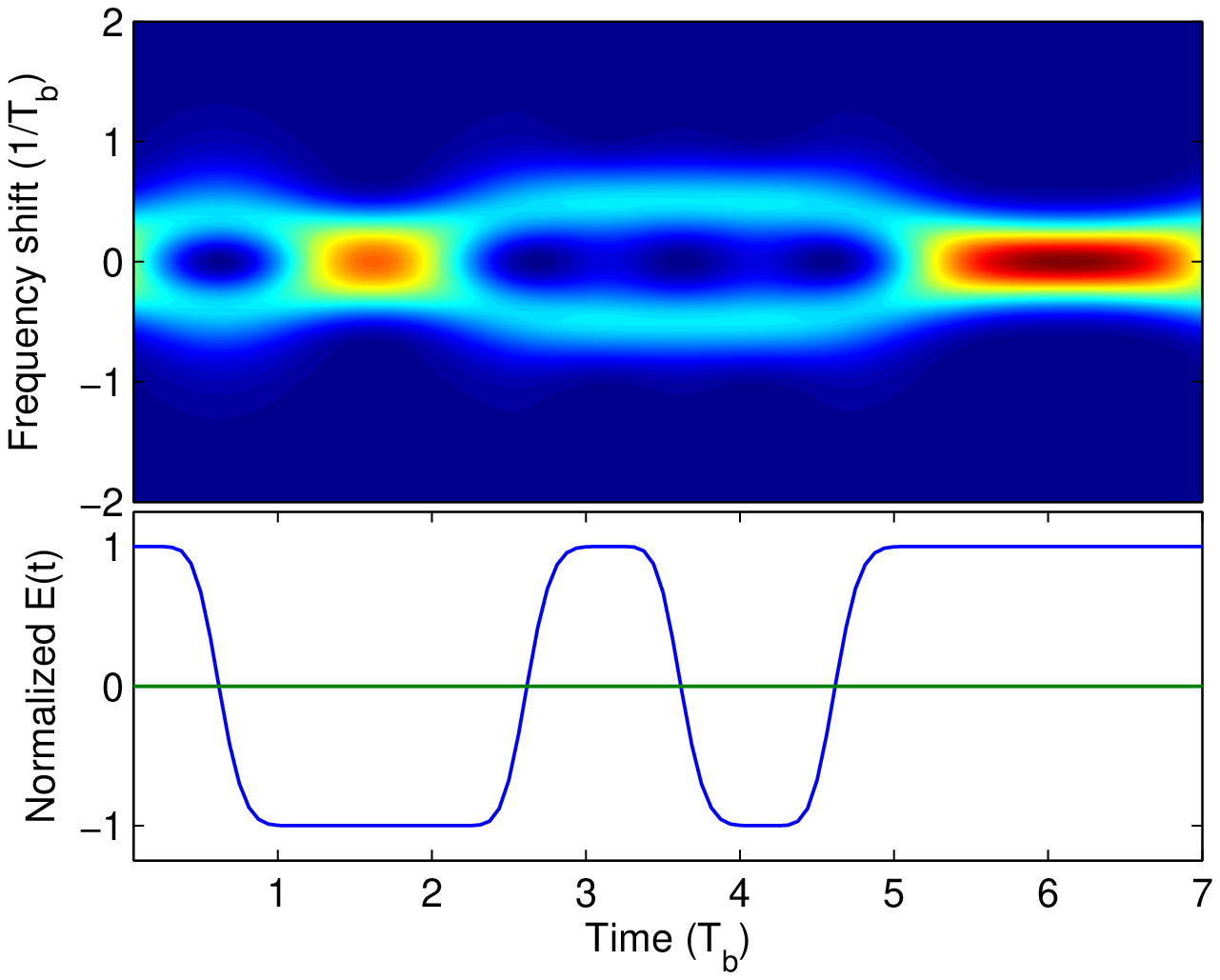} &
\includegraphics[width = 0.45 \textwidth]{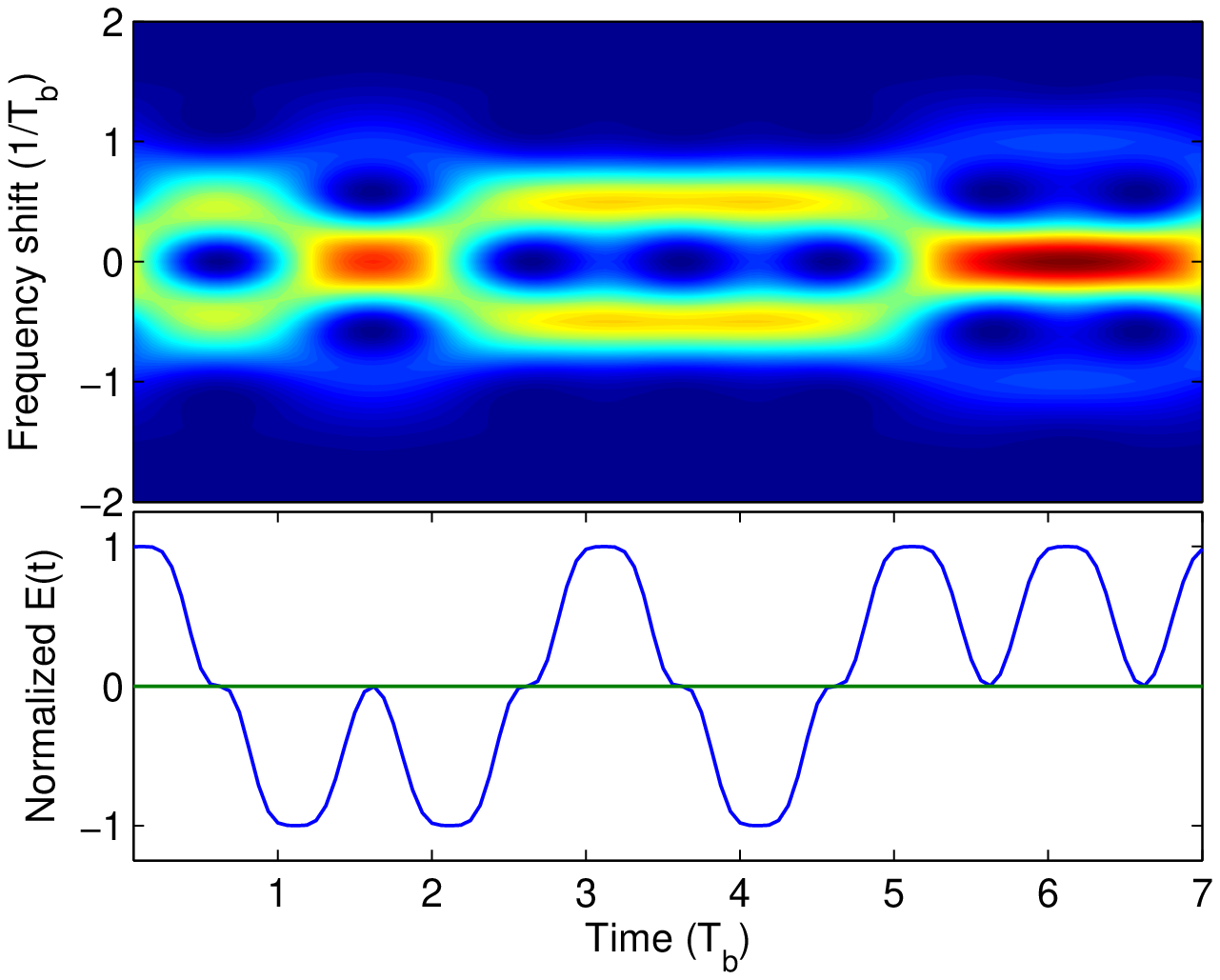} \\
(a) NRZ-DPSK & (c) RZ-DPSK \\
\includegraphics[width = 0.45 \textwidth]{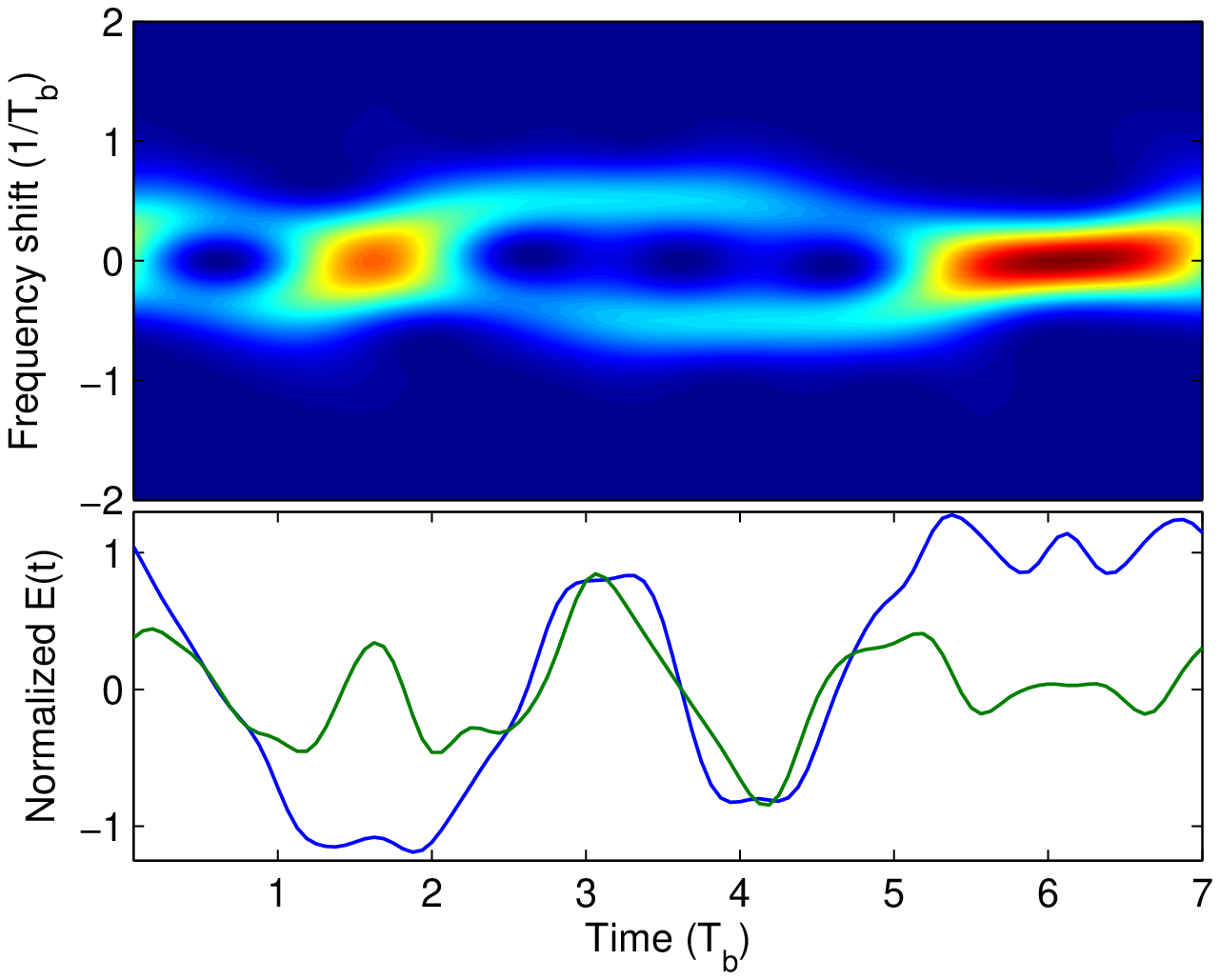} &
\includegraphics[width = 0.45 \textwidth]{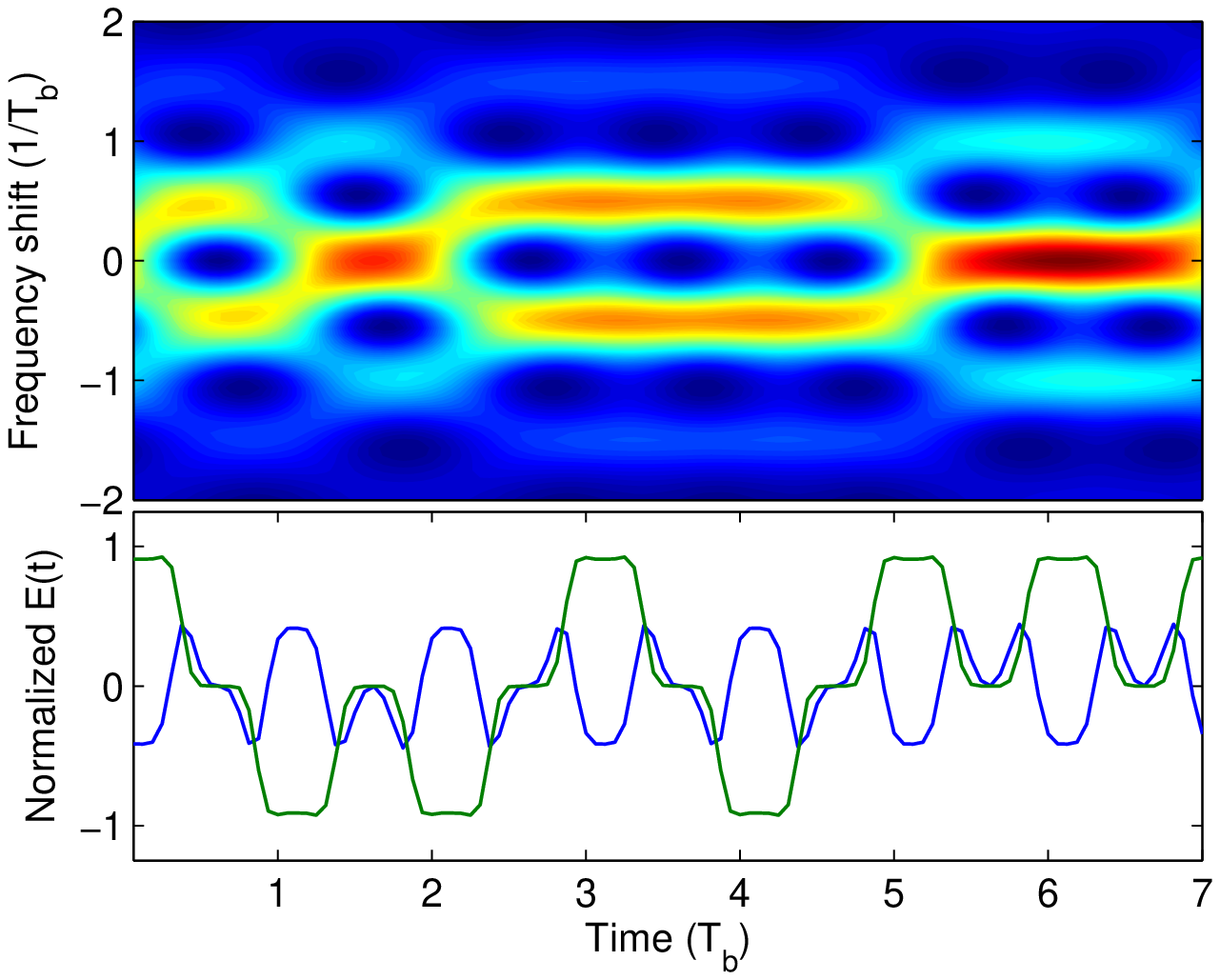} \\
(b) NRZ-DPSK with dispersion & (d) RZ-DPSK with SPM 
\end{tabular}
\end{center}
\caption{The TROF trace for 7-bit NRZ- and RZ-DPSK signals.
(a) NRZ-DPSK signal. (b) NRZ-DPSK signal with chromatic dispersion.
(c) RZ-DPSK signal. (d) RZ-DPSK signal with SPM.
(Blue lines: real part; Green lines: imaginary part)}
\label{figsim}
\end{figure}

\Fig{figsim} shows the simulated TROF traces for non-return-to-zero (NRZ) and return-to-zero (RZ) DPSK signals.
In \fig{figsim}, the optical filter of $H(\omega)$ has a Gaussian response and a full-width-half-maximum (FWHM) bandwidth of half the data rate of $0.5/T_b$, where $T_b$ is the bit period of the data stream.
\Fig{figsim} uses a 7-bit data pattern of $+1,-1,-1,+1,-1,+1,+1$ as an example for illustration purpose.
The center frequencies of the tunable filter [$\nu$ in \eqn{trofperiod}] tune between $\pm2/T_b$, i.e., twice the data rate.
The important properties of a TROF trace are all shown in \fig{figsim} for periodic phase-modulated signals.

\Fig{figsim}(a) is the TROF trace for NRZ-DPSK signal that is generated by a Mach-Zehnder amplitude modulator when the modulator is biased at the minimum transmission point and driven by a driving voltage swing of $2 V_\pi$ \cite{xu04, gnauck05}. 
From \fig{figsim}(a), there is an opening in the TROF trace, called a ``crater'', centered at a phase jump.
Consecutive phase jumps cascade to a large crater.
Without a phase jump, the TROF trace is concentrated around zero frequency, giving a ``ridge''.  
The craters and ridges of a TROF trace can be used to detect a DPSK signal, similar to the function of an asymmetric Mach-Zehnder interferometer \cite{gnauck02, xu04, gnauck05} or a frequency discriminator \cite{lyubomirsky05}.
Of course, the purpose here is to calculate both the phase and amplitude of a complex electric field.
The craters and ridges can be used to find the initial guess for the TROF algorithm explained in later parts of this paper.

\Fig{figsim}(b) shows the TROF trace of an NRZ-DPSK signal with fiber chromatic dispersion.
The amount of chromatic dispersion is equivalent to a 10-Gb/s signal propagated through 60-km of standard single-mode fiber with dispersion coefficient of $D = 17$ ps/km/nm.
The lower curves of \fig{figsim}(b) include the non-zero imaginary part of the electric field.
The real part of \fig{figsim}(b) is not as smooth as that for \fig{figsim}(a).
Similar to \fig{figsim}(a), even with fiber dispersion, the TROF trace of \fig{figsim}(b) has both craters and ridges.
Those craters and ridges can use to initiate the TROF algorithm.
Due to fiber dispersion, each crater and ridge in the TROF trace is rotated.
Later parts of this paper measure the TROF traces of NRZ-DPSK signals with and without chromatic dispersion and find the corresponding electric field of $E(t)$.

Not able to generate in our measurement, RZ-DPSK signal is the dominant signal format \cite{gnauck02, zhu03, rasmussen03, cai04, charlet04a, xu04, gnauck05}.
\Figs{figsim}(c) and (d) show the TROF traces of RZ-DPSK signal.
The RZ-DPSK signal uses the standard RZ pulses having a duty cycle of $1/2$ \cite{xu04, gnauck05}.
The major craters and ridges of the TROF trace of \fig{figsim}(c) are similar to that of \fig{figsim}(a).
The subtle difference is both thicker and higher crater ``rim'' for RZ- than NRZ-DPSK signals. 
\Fig{figsim}(c) also shows some satellite craters among both sides of a ridge, mostly due to the frequency sideband of RZ-DPSK signals.
Both the TROF trace of \figs{figsim}(a) and (c) are symmetric with respect to the signal frequency, showing a signal without frequency chirp.

RZ-DPSK signal with chromatic dispersion has a TROF trace largely the same as  that of \fig{figsim}(a).
\Fig{figsim}(d) shows the TROF trace of an RZ-DPSK signal with self-phase modulation (SPM).
The mean nonlinear phase shift is 1 rad.
Unlike the case with chromatic dispersion of \fig{figsim}(b), the signal spectrum is broadened for signal with SPM.
The craters among both sides of a ridge are larger and deeper than \fig{figsim}(c).
Both the craters and ridges are also slightly rotated.

The FWHM bandwidth of the filter is 50\% the data rate for the TROF traces of \fig{figsim}.
Qualitatively, if the filter bandwidth is too wide, the short craters between consecutive phase jumps may not able to cascade to a long crater.
If the filter bandwidth is too narrow, the short ridge may degenerate to a shallow crater.
Currently, there is no study on the optimal bandwidth to measure the TROF trace.
Practical measurement may use a bandwidth about 40 to 60\% the peak-to-notch bandwidth of the signal. 

\section{TROF Measurement and TROF Algorithm}
\label{sec:inv}

\begin{figure}[htbp]
\centering\includegraphics[width = 0.8 \textwidth]{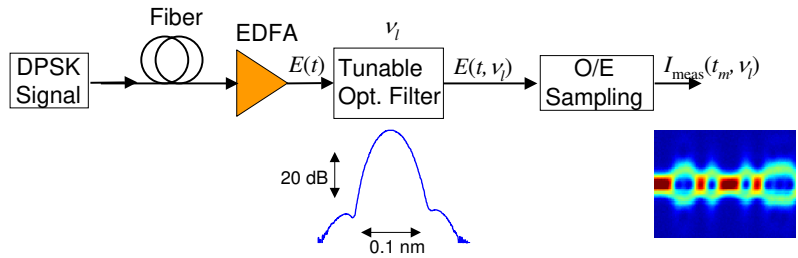}
\caption{Experimental setup to measure TROF trace}
\label{figsetup}
\end{figure}

The measurement of TROF traces is an experimental implementation for the expressions of \eqn{troftrace} or \eqn{trofperiod}. 
\Fig{figsetup} is the schematic diagram of the setup to measure TROF trace.
Operated at the wavelength of 1533 nm, a 10-Gb/s NRZ-DPSK signal is generated by the method described for \fig{figsim}(a).
For a practical communication signal, the signal is a $2^7-1$ PRBS with a period of $T = 12.7$ ns.
The PRBS includes all permutations of 7-bit length pattern except the all zero pattern. 
The NRZ-DPSK signal is passed to the TROF measurement equipment with or without passing through optical fiber.
The fiber has a dispersion coefficient of $D = 17$ ps/km/nm.
An Erbium-doped fiber amplifier (EDFA) is used to compensate for the loss at the tunable optical filter.

\begin{figure}[htbp]
\centering\includegraphics[width = \textwidth]{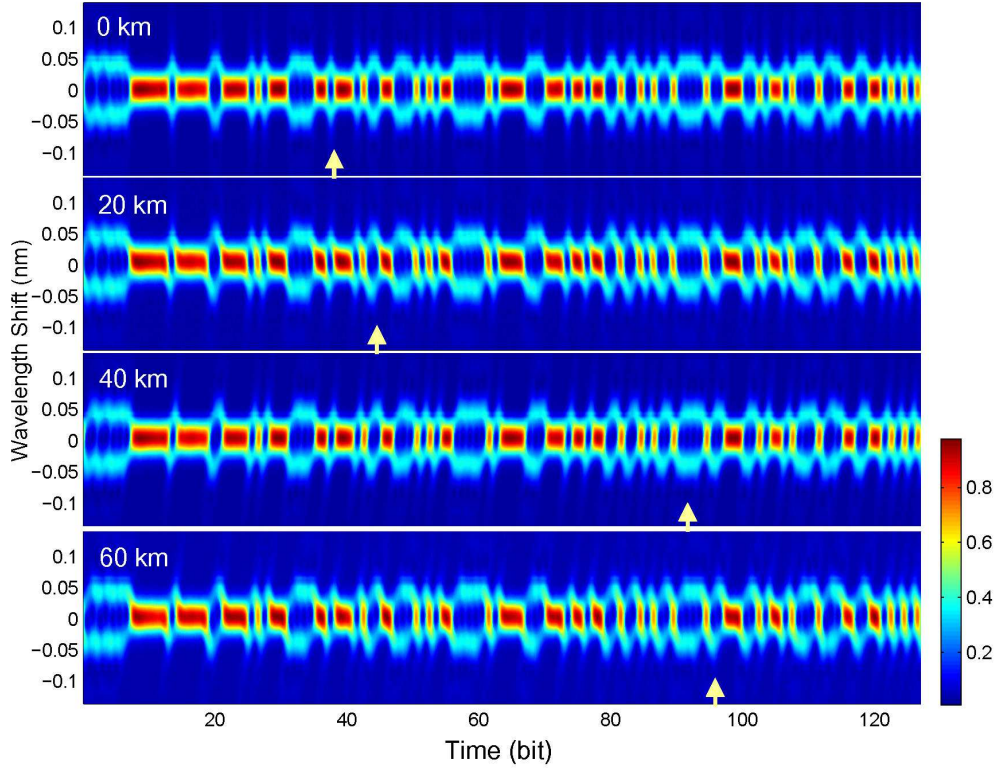}
\caption{Measured TROF traces for NRZ-DPSK signal after 0-, 20-, 40-, and 60-km of single-mode fiber.
The arrows are the beginning of data acquisition.}
\label{figsono}
\end{figure}

TROF equipment consists of a tunable optical filter, corresponding to the different response of $H(\omega - \nu_l)$ with center frequencies of $\nu_l$, and a high-speed optical-to-electric converter.
Similar to \cite{agilent}, \Fig{figsetup} uses the monochromator in Agilent 86146B with a FWHM bandwidth of 0.04 nm as the tunable optical filter.
Operated around the wavelength of 1533 nm, the FWHM bandwidth is about 5 GHz.
The optical-to-electrical converter is the sampling module of Agilent 86116A together with the 86100B digital communication analyzer mainframe.
The sampling module has a bandwidth of 53 GHz for optical signal, more than sufficient for 10-Gb/s signal.
From the physical properties of the monochromator \cite{wildnauer93}, the frequency response of $H(\omega)$ is linear phase without chirp.
The transfer function of $H(\omega)$ is also shown in \fig{figsetup}.
\Fig{figsono} shows the measured TROF traces for a NRZ-DPSK signal after the propagation of 0, 20, 40, and 60 km of standard single-mode optical fiber.

The TROF traces of \fig{figsono} have $N =2048$ evenly sample points.
With 64 centered frequency of $\nu_l$ scanned within $\pm 0.14$ nm of the signal wavelength, each TROF trace is a $64 \times 2048$ array of data. 
In the wavelength of 1533 nm, $\pm 0.14$ nm corresponds to $\pm 17.5$ GHz.
Unlike FROG or similar technique with a square data array and a fast algorithm \cite{kane99, reid99}, only the conventional method of \cite{trebino93} is applicable for the rectangular array for TROF. 

Due to propagation delay, the TROF traces are acquired with different delay.
The arrow of each TROF trace indicates the beginning of data acquisition.
With an acquisition window the same as the data period, the TROF traces of \fig{figsono} are aligned by post-processing and normalized with peak unity intensity.

Other than a longer pattern with $2^7-1$ bits, the TROF traces of \fig{figsono} are very similar to \figs{figsim}(a) and (b) without and with chromatic dispersion, respectively.
Without chromatic dispersion, similar craters and ridges appear in \fig{figsono} symmetrical with respect to the signal wavelength.
With the increases of chromatic dispersion, the TROF trace becomes more asymmetric with respect to the center wavelength and clockwise rotated. 
The rotation angle increases with the amount of chromatic dispersion.
The TROF trace of \fig{figsim}(b) is the vertical flip of  those in \fig{figsono} as the $y$-axis is frequency in \fig{figsim}(b) and the $y$-axis is wavelength in \fig{figsono}.
Because the increase of wavelength decreases the frequency, the TROF trace with wavelength shift as $y$-axis is the vertical flip of the TROF trace with frequency shift as $y$-axis.

From the Appendix, the periodic electric field of $E(t)$ is uniquely determined by its TROF trace up to a constant factor.
However, the method from the Appendix cannot convert to a practical numerical method because noise in the TROF trace leads to divergent electric field of $E(t)$.

With a measured TROF trace of $I_\mathrm{meas}(t_m, \nu_l)$ at each time sample of $t_m$ of \fig{figsono}, similar to the method of \cite{trebino93},  numerical optimization is used to find the complex electric field of $E(t)$ from \eqn{ESignal}.
The TROF algorithm minimizes the mean-square error (MSE) of 
\begin{equation}
\mathcal{E} = \sum_l \sum_m \left[I_\mathrm{meas}(t_m, \nu_l) - I_\mathrm{TROF}(t_m, \nu_l) \right]^2,
\label{MSerror}
\end{equation} 
where $I_\mathrm{TROF}(t_m, \nu_l)$ is calculated numerically using \eqn{trofperiod}, similar to the traces of \fig{figsim}.

There are many numerical optimization methods to find the values of $E_m = E(t_m)$ to minimize the MSE of \eqn{MSerror} \cite{recipe}.
Conjugate gradient method is especially suitable for this optimization problem.The gradient of the MSE, $\nabla \mathcal{E}$, composites by the differentiation of
\begin{equation}
\frac{\partial \mathcal{E}}
{\partial E_k}
   =  -2 \sum_l \sum_m  \left[I_\mathrm{meas}(t_m, \nu_l) - I_\mathrm{TROF}(t_m, \nu_l) \right] \frac{\partial I_\mathrm{TROF}(t_m, \nu_l)}
{\partial E_k}.
\label{partialE}
\end{equation}
If the TROF trace has $N$ evenly samples in time domain, based on discrete Fourier transform, the Fourier coefficients of $c_k$ are $c_k = \frac{1}{N} \sum_m E_m \exp\left(- j 2 \pi k m/N \right)$, and
\begin{equation}
\frac{\partial I_\mathrm{TROF}(t_m, \nu_l)}
{\partial E_k} = 
 \frac{2}{N} E(t_m, \nu_l) \sum_{k_1} H \left( \frac{2 \pi k_1}{T} - \nu_l \right) \exp \left[ \frac{2 \pi j k_1 (m - k)}{N} \right], 
\label{partialI}
\end{equation}
where  $E(t_m, \nu_l)$ is the output electric field of optical filter (see \fig{figsetup}), given by
\begin{equation}
E(t_m, \nu_l) =  \sum_{k} c_{k_1} H \left( \frac{2 \pi k}{T} - \nu_l \right) \exp \left( \frac{2 \pi j k m}{N} \right).
\label{Eoutput}
\end{equation} 
Other than $E(t_m, \nu_l)$, the term of \eqn{partialI} depends solely on the difference of $m - k$.
Numerically, the values of \eqn{partialI} and the electric field of \eqn{Eoutput} can be evaluated by fast Fourier transform.

\begin{figure}[htbp]
\begin{center}
\includegraphics[width = 0.95 \textwidth, height = 0.35 \textwidth]{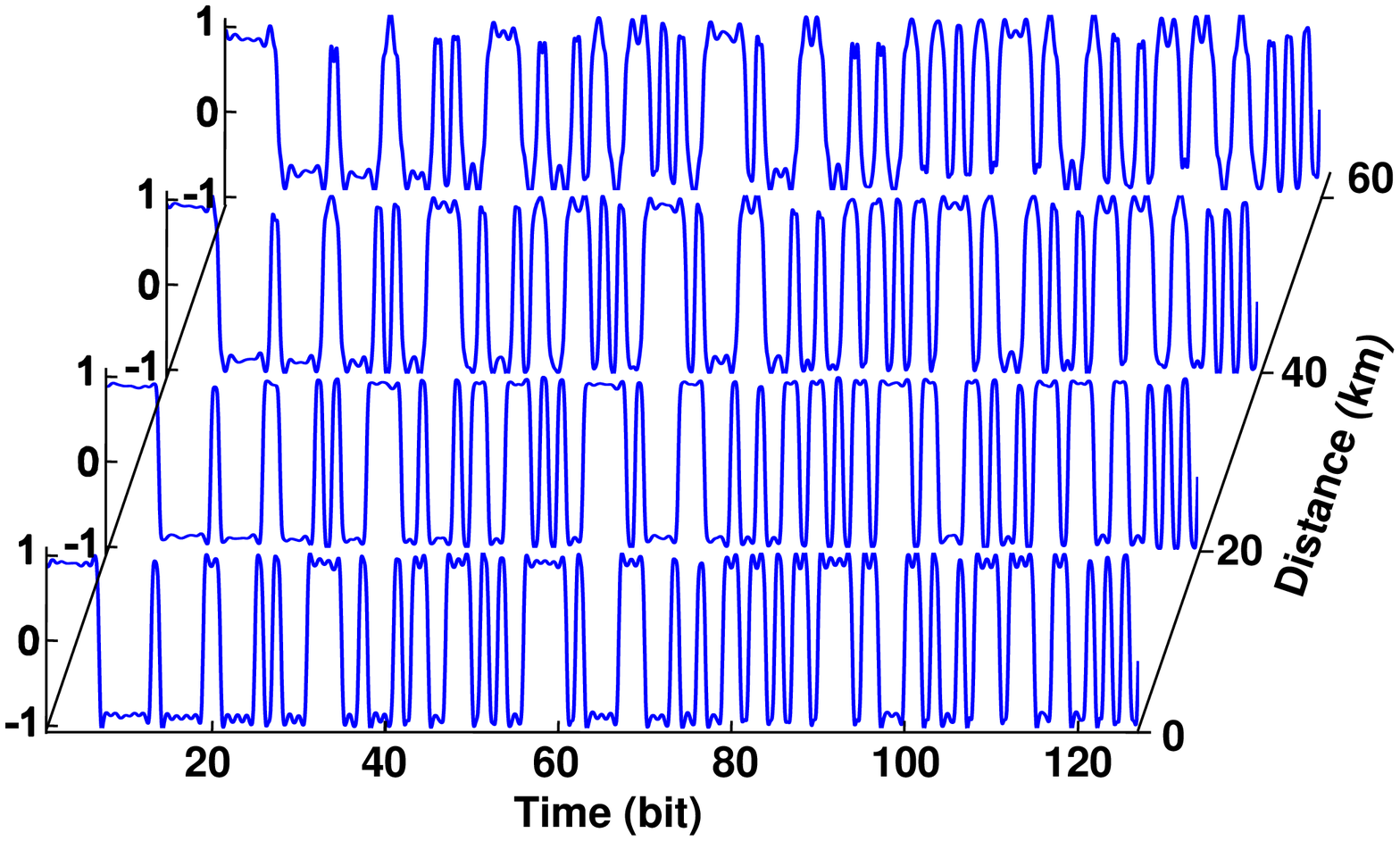} \\
(a) $\Re\left\{E(t)\right\}$\\
\includegraphics[width = 0.95 \textwidth, height = 0.35 \textwidth]{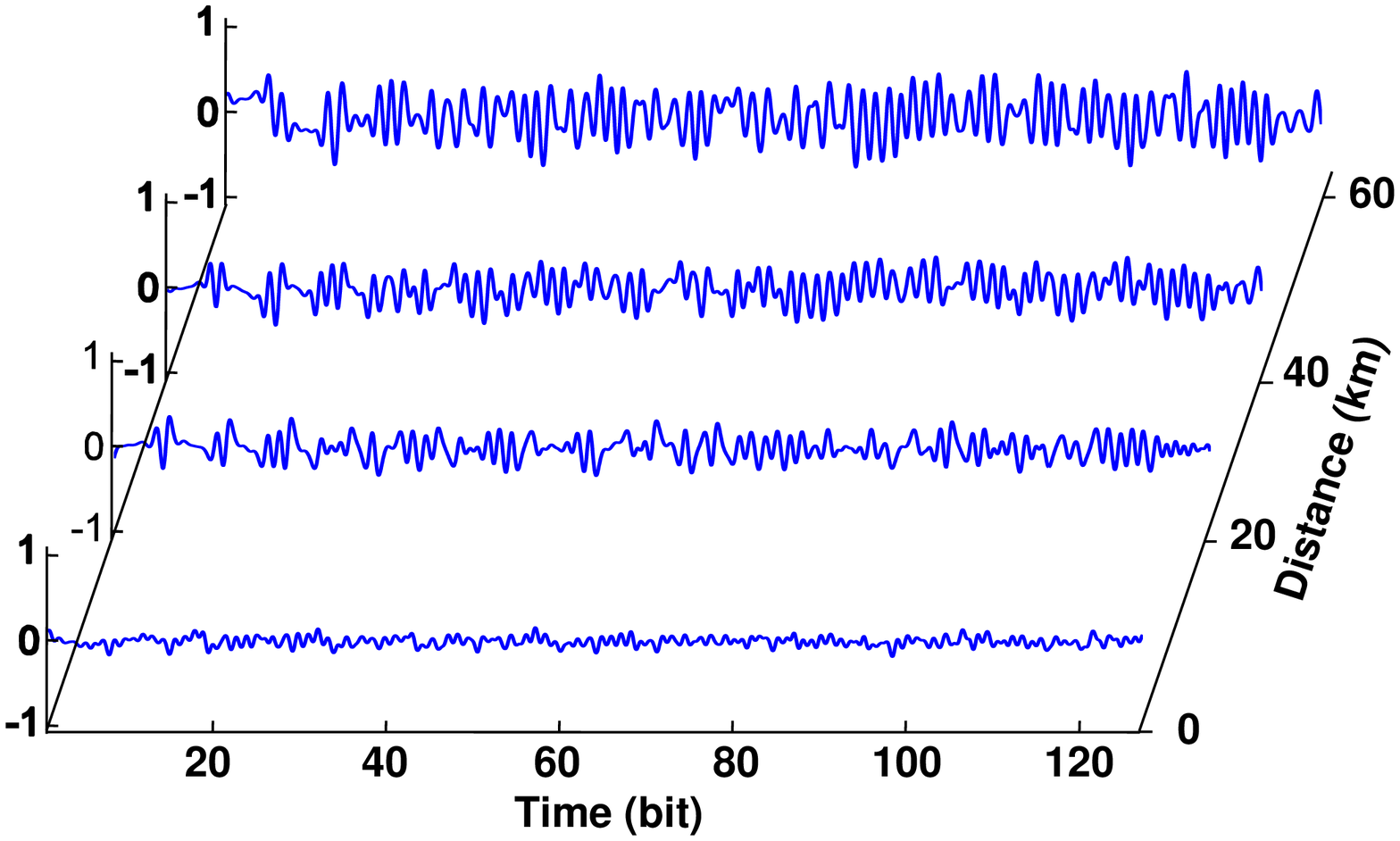} \\
(b) $\Im\left\{E(t)\right\}$
\end{center}
\caption{The normalized complex electric field calculated numerically from the TROF traces of \fig{figsono}. (a) Real part, (b) imaginary part.}
\label{figefield}
\end{figure}

\Fig{figefield} shows the complex-value electric field calculated by the TROF algorithm initiated by the craters and ridges of the sonogram.
Even with various amount of chromatic dispersion, the TROF algorithm is initiated by the same waveform.
The TROF algorithm converges within less than 20 iterations to a MSE about 0.5\%, 1.3\%, 0.9\%, and 1.1\% for 0-, 20-, 40-, and 60-km of single-mode fiber, respectively.
The real part of the waveform of \fig{figefield}(a) has distortion increased with chromatic dispersion.
The imaginary part of the waveform of \fig{figefield}(b) is originally very close to zero without fiber dispersion but becomes larger with the increase of chromatic dispersion.

In general and without special precaution, the MSE increases with the fiber distance mainly due to timing error.
Over the measurement interval, the fiber may be expanded or contracted with environment temperature.   
The small difference of propagation distance induces timing error between the value of $I_\mathrm{meas}(t, \nu_l)$ in early and later parts of the measurement.
The TROF trace measurement of \fig{figsono} already shortens the measurement time to minimize this effect.
With chromatic dispersion, there is also bigger difference between the initial guess and the optimized electric field than the case without chromatic dispersion.
With longer distance, the amplifier noise for 60-km measurement is also larger than that without optical fiber.

\section{Discussion}
\label{sec:dis}

From the theory of \cite{kikuchi01}, the accuracy of sonogram is limited by the time-resolved power of the electrical sampling head.
A resolution down to femtosecond is required to characterize optical short pulses \cite{chilla91, taira01, reid99}.

Requiring a resolution of few picoseconds, the TROF traces of \fig{figsono} are measured by an electrical sampling head with 53 GHz of  bandwidth.
Equivalently, other than noise, the measured TROF traces are $I_\mathrm{meas}(t, \nu_l) = I_\mathrm{TROF}(t, \nu) \otimes h_s(t)$, where $\otimes$ denotes convolution and $h_s(t)$ is the impulse response of the optical sampling head.  
To include the contribution of $h_s(t)$ to the theory of the Appendix, $i_m(\nu)$ defined by \eqn{imtheory} becomes $i_m(\nu)H_s \left( 2 \pi m/T \right)$, where $H_s(\omega)$ is the frequency response of the optical sampling head.
For 10-Gb/s signal, only the parts of $H_s(\omega)$ with frequencies less than 20 GHz are important.
With a 53-GHz sampling head, $h_s(t)$ is a very short impulse response and $H_s(\omega)$ has a very wide bandwidth.
Without chromatic dispersion, the ripples of \fig{figefield} may be from the electrical sampling head.
Measured using the same sampling head, the intensity of an optical signal has similar ripples.
In the measurement of optical signal for communication purpose, the ripples from the sampling head are usually tolerated as measurement artifacts.

In this paper, periodic signal is considered for the convenience of using fast Fourier transform.
The number of data samples is a power of 2 for fast Fourier transform.
With proper windowing, the algorithm can be generalized to non-periodic signal.

In previous section, the TROF algorithm is developed assuming that $E_m$ is a real signal.
For practical complex signal of $E_m$, the number of unknown variables is doubled.
Only minor modifications are required for both \eqn{partialE} and \eqn{partialI} for complex signal.

The craters and ridges of the spectrogram are special for phase-modulated signals.
For other signal types without a proper initial guess, the TROF algorithm requires time-consuming calculation.
The TROF algorithm may have ``stagnation'' problem similar to FROG algorithm \cite{trebino97, kane99, trebino}. 

\section{Conclusion}
\label{sec:end}

The time-frequency dual of FROG, TROF is used for periodic phase-modulated optical signals.
TROF uses only electrical sampling followed an optical-to-electrical converter.
To our knowledge, this is the first application of similar techniques to signal with infinite support instead of short pulse with finite support.

The uniqueness of the signal from a TROF trace is proved analytically for periodic signal.
Based on standard setup for time-resolved chirp measurement, the TROF traces are measured for phase-modulated signals with and without chromatic dispersion.
The signal is determined from the measured TROF  trace based on numerical optimization to minimize the difference between the analytical and measured TROF traces.
The TROF algorithm converges within 20 iterations to a MSE around or less than 1\%.

\section*{Appendix: Theory of TROF for Periodic Signals}

The sonogram or spectrogram of \eqn{trofperiod} can be expressed as
\begin{equation}
I_\mathrm{TROF}(t, \nu)
 =  \sum_k \sum_m
    c_k c_{k-m}^*
    H\left( \frac{2 \pi k}{T} - \nu \right) H^*\left[ \frac{2 \pi (k-m)}{T} - \nu \right] \exp \left(\frac{2 \pi j m t}{T} \right).
\end{equation}
Defined $h_m(\tau)$ and $r_m(\tau)$ as
\begin{eqnarray}
h_m(\tau) &=& \frac{1}{2 \pi} \int H(\omega) H^* \left( \omega - \frac{2 \pi m}{T} \right)
                e^{j \omega \tau} \ud \omega, \\
r_m(\tau) &=& \sum_k c_k c^*_{k-m} e^{-2 \pi j k \tau/T}, 
\end{eqnarray} 
we obtain
\begin{equation}
i_m(\nu)
 = \int h_m(\tau) r_m(\tau)  e^{j \nu \tau} \ud \tau ,
\end{equation}
\noindent where 
\begin{equation}
i_m(\nu) = \frac{1}{T} \int_{0}^T I_\mathrm{TROF}(t, \nu) 
\exp \left(-\frac{2 \pi j m t}{T} \right) \ud t.
\label{imtheory}
\end{equation}    

Mathematically, with some algebra, we obtain
\begin{equation}
r_m(\tau) = \frac{1}{2 \pi h_m(\tau)} \int i_m(\nu) e^{-j \nu \tau} \ud \nu.
\end{equation}
and
\begin{eqnarray}
E(t) &=& \sum_k   c_k  \exp \left( 2 \pi j k t/T \right) \nonumber \\
   & = & \frac{1}{ 2 \pi T c_0^*} \int_{0}^{T}  \sum_k r_k(-\tau) \exp \left[ 2 \pi j k (t+ \tau)/T \right] \ud \tau  \\
   & = & \frac{ 1}{2 \pi T c_1^*} \int_{0}^{T}  \sum_k r_{k-1}(-\tau) \exp \left\{ 2 \pi j [(k -1 )t+ k \tau]/T \right\} \ud \tau. 
\label{Etheory}
\end{eqnarray}

The above analysis is similar to that in \cite{cohen89, kikuchi01} for signal with finite support.
The sonogram is basically the time-frequency distribution of a periodic signal with a kernel of $h_m(\tau)$.
Although similar theory is not discussed in both \cite{cohen89, kikuchi01}, the principle remains the same.

The above analysis confirms the existence and the uniqueness of the signal of $E(t)$ from the sonogram of $I_\mathrm{TROF}(t, \nu)$ up to a constant factor of $c_0$ or $c_1$.
However, the kernel of $h_m(\tau)$ is very small for large $\tau$.
For example, if $H(\omega) = \exp(-\omega^2/2\omega_0^2)$ as a Gaussian filter,
the kernel of
\begin{equation}
h_m(\tau) \propto
   \exp \left(
      - \frac{\pi^2 m^2}{ \omega_0^2 T^2}
      + \frac{ \pi j m \tau}{ T}
       - \frac{\tau^2 \omega_0^2}{4}
   \right)
\end{equation}
becomes very small for $\tau \gg 1/\omega_0$.
In theory, $\int i_m(\nu) e^{-j \nu \tau} \ud \nu$ is either comparable or smaller than $h_m(\tau)$.
In practical calculation, a small $h_m(\tau)$ is numerically difficult to handle.
If the measurement of the TROF trace of $I_\mathrm{TROF}(t, \nu)$ has noise or the calculation of $i_m(\nu)$ has small numerical error, the electric field of $E(t)$ has enormous error.
Nevertheless, the technique is applicable for short pulse with finite support \cite{taira01}.

\end{document}